\documentclass[twocolumn,prl,aps,showpacs]{revtex4}
\usepackage{amssymb}
\usepackage{epsfig}
\usepackage{amsbsy}
\usepackage{amsmath}
\usepackage{graphicx}

\begin{document}

\title{Simulating Wess-Zumino Supersymmetry Model in Optical Lattices}

\author{Yue Yu$^1$ and Kun Yang$^2$}
\affiliation{$^1$Institute of Theoretical Physics, Chinese Academy
of Sciences, P.O. Box 2735, Beijing 100190, China\\
$^2$NHMFL and
Department of Physics, Florida State University, Tallahassee,
Florida 32306, USA}

\date{\today}

\begin{abstract}{\small
We study a cold atom-molecule mixture in two-dimensional optical
lattices. We show that by fine-tuning the atomic and molecular
interactions, Wess-Zumino supersymmetry (SUSY) model in
2+1-dimensions
 emerges in the
low-energy limit and can be simulated in such mixtures. At zero
temperature, SUSY is not spontaneously broken, which implies
identical relativistic dispersions of the atom and its superpartner,
bosonic diatom molecule. This defining signature of SUSY can be
probed by single-particle spectroscopies. Thermal breaking of SUSY
at a finite temperature is accompanied by a thermal Goldstone
fermion, i.e., phonino excitation. This and other signatures of
broken SUSY can also be probed experimentally.}

\end{abstract}
\pacs{67.85.Pq, 37.10.Jk,
11.30.Pb~~~~~~~~~~~~~~~~~~http://dx.doi.org/10.1103/PhysRevLett.105.150605}
 \maketitle

 {\it Introduction. --} Wess and Zumino proposed the
first space-time supersymmetry (SUSY) model (WZ-SUSY model) 36 years
ago \cite{wz}. Since then SUSY has become a fundamental ingredient
of theories beyond the standard model in high-energy physics
\cite{Weinberg}. However, none of super partners of the known
elementary particles have been found thus far; it remains to be seen
if they can be detected in the energy range of Large Hadron
Collider.

On a different front, {\em nonrelativistic} SUSY (a Bose-Fermi
symmetry unrelated to space-time symmetry) has attracted
considerable recent interest in the cold atom community, as it can
be realized by using Bose-Fermi atom (molecule) mixtures which are
loaded in optical lattices. Examples include attempts to simulate
the nonrelativistic limit of superstring by trapping fermionic atoms
in the core of vortices in a Bose-Einstein condensate \cite{string};
study of the SUSY effect in an exactly solvable one-dimensional
Bose-Fermi mixture with Bethe ansatz \cite{id}; and SUSY models for
nonrelativistic particles in various dimensions\cite{Lo,YY,SYS}. In
Ref. \cite{YY}, we studied perhaps the simplest cold atom SUSY model
and discussed detecting the Goldstino-like mode due to SUSY breaking
by measuring a single fermion spectral function. In a further work
\cite{SYS}, we developed a SUSY response theory to photoassociation
in a cold fermionic atom system. Although these studies are
interesting and some results may be general for many SUSY systems
[7], SUSY in these nonrelativistic systems is very different from
the relativistic (or space-time) SUSY in high-energy physics.

In this Letter, we propose a way to simulate the simplest {\em
relativistic} SUSY model, the WZ-SUSY model \cite{wz}. We show that
it can emerge in the low-energy limit of a cold atom-molecule
mixture in properly chosen two-dimensional lattices. The first
requirement is the existence of Dirac points in the Brillouin zone.
Recently, such models based on a honeycomb lattice or graphenelike
structure have been proposed \cite{optlatt}. In these models, two
Dirac points $K$ and $K'$ are related to each other by $K'=-K$, as
required by time-reversal symmetry. This means This means that two
fermionic atoms which form a usual BCS pair or a diatom molecule
belong to two different Dirac points. To simulate the WZSUSY model,
however, one needs a Klein-Gordon field as the Dirac fermion¡¯s
superpartner which corresponds to a diatom molecule made by two
Dirac fermions from the {\em same} Dirac point. Such molecules carry
a $2K\ne 0$ momentum and are energetically unfavorable as a result.
In a recent work, Lee attempted to avoid this difficulty by
introducing frustrated hopping for the molecules, such that the
boson dispersion has minima at $\pm 2K$ instead of zero \cite{Lee}.
It is found that the massless WZ-SUSY model emerges at the boson's
superfluid-insulator critical point.

In this work, we show that the WZ-SUSY model can emerge not only at
the critical point. We use a lattice  model studied recently by Liu
{\em et al.} \cite{LLWS} instead. This is a square lattice model in
which the Dirac points at $K=(0,0)$ and $K'=(0,\pi)$ are their own
negatives, as $(0,-\pi)\equiv(0,\pi)$. This means a diatom molecule
made of two atoms from the {\em same} Dirac points has zero-momentum
for $2K=(0,0)$ and $2K'=(0,2\pi)\equiv (0,0)$. With this setup, we
can simulate the WZ-SUSY model more straightforwardly, after
appropriate interactions are introduced and fine-tuned.

 The research interest in WZ-SUSY models has been renewed recently\cite{rsymm}.
No spontaneous breaking of the SUSY implies there are equal poles in
the single-particle spectral functions of both the Dirac field and
the Klein-Gordon field. A further calculation showed that these
single-particle spectral functions are not renormalized from their
free particle ones \cite{phonino}. This is the identifier of the
SUSY and may be detected by the established techniques of the
single-particle spectroscopies \cite{sps}. It is known that a
thermal bath always breaks the SUSY \cite{tb} and this thermal
breaking of the SUSY is accompanied by a thermal Goldstone fermion,
phonino \cite{phonino}; thus studying this model at finite
temperature sheds light on physics of SUSY breaking.

There are many studies of SUSY in space-time lattice
models\cite{stlsusy}. The significant difference between the the
present work and those lattice SUSY models is that while the latter
are supersymmetric on the lattices, we study the {\em emergence} of
SUSY from a microscopic space lattice (but continuous time) model
with no SUSY to begin with.

{\it Free Fermion Lattice Model and Continuum Limit. --} We briefly
recall the lattice model proposed in Ref. \onlinecite{LLWS}.
Consider a single-component fermionic atom gas loaded in a square
lattice. The potential minimum in the sublattice $A$ is higher than
that in the sublattice $B$. Two states with the energy difference
$2M$, the $s$-orbital at the $A$-sites and the $p$-orbital at the
$B$-sites, form a pseudospin-1/2 subspace. The sublattices are
anisotropic with $1,2,3,$ and 4, the next nearest neighbor sites
(Fig. 1(b)). The hoppings between the nearest and next nearest sites
are taken into account. The corresponding hopping amplitudes are
$t_{A,A+\delta_{x(y)}}=-t_{A,A-\delta_{x(y)}}=t_{AB}$,
$t_{A1}=t_{A3}$, $t_{A2}=t_{A4}$, $t_{B1}=t_{B3}$, and
$t_{B2}=t_{B4}$ with $\delta_{x}$ ( $\delta_{y}$) being the unit
vector in the $x$- ($y$-) direction. In addition, a periodic gauge
field generated by two opposite-traveling standing wave laser beams
coupling with atoms\cite{spielman} is introduced. This gives rise to
a tunable staggered Peierls phase $\pm\theta_0$ along the vertical
links  and vanishing in the horizontal and $1,2,3,4$ links. With
these the single-fermion Hamiltonian is given by
\begin{eqnarray}
H({\bf k})=p_x({\bf k})\sigma_x+p_y({\bf k})\sigma_y+h_z({\bf
k})\sigma_z,
\end{eqnarray}
where $p_x=2t_{AB}\sin\theta_0\sin(k_ya)$,
$p_y=2t_{AB}(\sin(k_xa)+\cos\theta_0\sin(k_ya))$ and
$h_z=-M-t_0\cos(k_xa)\cos(k_ya)-2\tilde t\sin(k_xa)\sin(k_ya)$ with
$t_0=t_{A1}-t_{B1}+t_{A2}-t_{B2}$ and $\tilde
t=(t_{A1}-t_{B1}+t_{B2}-t_{A2})/2$. When $M=\pm t_0 \ne 0$, there is
a unique gapless Dirac point: either $K=(0,0)$ or $K'=(0,\pi)$. We
choose $\theta_0=\pi/2$ and define the 'speed of light'
$v_s=2t_{AB}a$. In the continuum limit and near the Dirac points,
$p_x(K+\delta k)\sim 2t_{AB}a\delta k_y\equiv v_s q_x$ and
$p_x(K'+\delta k)\sim -2t_{AB}a\delta k_y=-v_s q_x$ ; $p_y(K+\delta
k)\sim 2t_{AB}a\delta k_x\equiv v_sq_y$ and $p_y(K'+\delta k)\sim
2t_{AB}a\delta k_x\equiv v_sq_y$; and $h_z(K+\delta k)=-M-t_0=m_0$
and $h_z(K'+\delta k)=-M+t_0=m_\pi$. Thus, for the Dirac fields
$\xi(x)$ near $K$ and $\zeta(x)$ near $K'$, the effective
Hamiltonian reads
\begin{eqnarray}
H_c^{(0)}&=&v_s\int
d^2x\xi^\dag(-i\alpha_+^a\partial_a+m_0\sigma_z)\xi\nonumber\\
&+&v_s\int
d^2x\zeta^\dag(-i\alpha_-^a\partial_a+m_\pi\sigma_z)\zeta,\label{free}
\end{eqnarray}
where $\alpha_\pm^1=\pm\sigma^x$ and $\alpha_\pm^2=\sigma^y$; the
two-component Dirac field $\xi({\bf r})$ is given by
\begin{figure}[htb]
\begin{center}
\includegraphics[width=7cm]{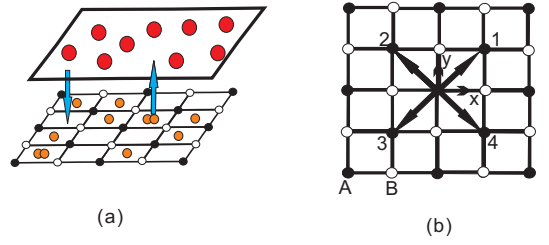}
\end{center}
\vspace{-0.5cm}
 \caption{\label{fig:Fig.1}
(a) Josephson tunneling between the atom-molecule mixture (lower
lattice plane) and the dimolecule Bose-Einstein condensate nearby
(upper plane). The orange dots are molecules in the mixture and red
dots are dimolecules. Fermionic atoms are in the lattice sites. (b)
The square lattice where 1,2,3,4 denote the next nearest neighbor
sites. }
\end{figure}
$$
\left(\begin{array}{c}\xi_1\\
\xi_2\end{array}\right)=\frac{1}{\sqrt 2}\int_0^\Lambda d^2 q
e^{i{\bf q\cdot x}}\biggl[\xi_{1 \bf q}\left(\begin{array}{c}
\theta_{\bf q}\\1\end{array}\right) +\xi^\dag_{2,-\bf q}
\left(\begin{array}{c} \theta_{\bf q}\\-1\end{array}\right)\biggr]$$
with $\theta_{\bf q}=\frac{q_x+iq_y}{|\bf q|}=-\theta_{-\bf q}$ and
similarly for $\zeta$. The momentum cut-off $\Lambda$ corresponds to
that when lattice fermion dispersion deviates severely from the
linear one. The mass terms here can be fine-tuned. When $M=\pm t_0$
which is not zero in this lattice setup\cite{LLWS}, one of the Dirac
field is massless and another is massive. The latter can be
integrated out in the low-energy limit. We note the zero matter
density (necessary for Lorentz invariance) in relativistic quantum
field theory corresponds to fermionic atoms being at half filling in
this lattice realization. After an external source is introduced,
fermion number (including those forming molecules) will fluctuate
but average at half filling. To facilitate pairing or molecule
formation, we introduce attraction between fermionic atoms, which is
modeled in a two-channel fashion below. For such spinless fermions,
the two-atom attraction and $p$-wave type bound state have already
been achieved experimentally\cite{pexp}.

{\it Two-channel Model. --}  We take $m_0=0$ and $m_\pi\ne 0$, and
integrate out $\zeta$ in the low-energy limit where only the states
with their energy lower than min$\{m_\pi,E_\Lambda\}$ are relevant.
We now extend our Hamiltonian to a two-channel model, i.e., the
lowest two hyperfine atom states with two-atom scattering states in
open channel and the two-atom bound state (Feshbach molecule) in
closed channel. We denote $\xi^{(o)}(x)$ the Dirac fermions in the
open channel and $\xi^{(c)}(x)$ the Dirac fermion in the closed
channel. Analogous to the many body theory of the atom-molecule
coherence in Ref. \onlinecite{DS}, the effective Lagrangian
describing this two-channel Dirac fermion model is given by
\begin{eqnarray}
&&{\cal L}= -\xi^{(o)\dag}\sigma^\mu\partial_\mu\xi^{(o)}\nonumber
-\xi^{(c)\dag}\sigma^\mu\partial_\mu\xi^{(c)}\\
&&+U^{(c)}\xi^{(c)\dag}_2\xi^{(c)\dag}_1\xi^{(c)}_1\xi^{(c)}_2
+U^{(co)}\xi^{(c)\dag}_2\xi^{(c)\dag}_1\xi^{(o)}_1\xi^{(o)}_2+h.c.,\nonumber
\end{eqnarray}
where $\sigma_\mu=(I,\sigma_x,\sigma_y)$ and
$\partial_\mu=(\partial_t,v_s\nabla)$. $U^{(c)}$ and $U^{(co)}$ are
the interaction between closed channel fermions and the interchannel
interaction, respectively. We have neglected the background
interaction in open channel. By introducing the pairing field
$\Delta({\bf r},t)$ for $\xi^{(c)}$ via a Hubbard-Stratonovich
transformation and integrating out $\xi^{(c)}$, the resulting
Lagrangian is given by
\begin{eqnarray}
{\cal L}(\xi^{(o)},\Delta)
=-\frac{1}2\xi^{(o)\dag}\sigma^y\sigma^\mu\partial_\mu\xi^{(o)}
-\frac{|\Delta|^2}{U^{(c)}}+{\rm Tr}\ln
G^{{(c)}-1}.\nonumber\\\label{el}
\end{eqnarray}
The inverse  of the propagator $G^{(c)}$ of $\xi^{(c)}$ is given by
$$
G^{(c)-1}=\left(\begin{array}{cc}
                0&i\sigma^\mu\partial_\mu\\
                -i\sigma^\mu\partial_\mu&0\end{array}\right)
-\left(\begin{array}{cccc}
                \Xi&0\\
                0&\Xi^\dag\\
                \end{array}\right)\nonumber
$$
where $\Xi=\Delta+U^{(co)}\xi_1^{(o)}\xi_2^{(o)}$. Expanding the
Lagrangian in powers of $\Delta$ and its gradients yields
\begin{eqnarray}
{\cal L}[\varphi] =-\frac{1}2\partial_\mu \varphi^\dag\partial^\mu
\varphi-\frac{1}2\varepsilon_m|\varphi|^2-\frac{\lambda}8|\varphi|^4+O(|\varphi|^6)
,\nonumber\\\label{kg}
\end{eqnarray}
where $\varphi\propto \Delta/U^{(c)}$ is the Feshbach molecular
field with the detuning energy $\varepsilon_m$ and the interacting
strength $\lambda\propto (U^{(c)})^2$. We have $v_b=v_s$ in the weak
coupling limit (i.e., $U^{(c,co)}$ much smaller all other energy
scales in the system including $m_\pi$ and $E_\Lambda$) due to
(emergent) Lorentz invariance. Lattice effects (which break Lorentz
invariance) give rise to nonuniversal corrections to $v_s$ and
$v_b$; thus tuning of one parameter ({\em e.g.}, molecule dispersion
through an additional lattice potential seen by the molecule only)
is needed to ensure $v_b=v_s$ to maintain Lorentz invariance in the
low-energy limit. Also included in (\ref{el}) is the Yukawa coupling
between $\varphi$ and $\xi^{(o)}$, {\it i.e.},  $ {\cal
L}_{\varphi\xi}=-\frac{g}2(\varphi\xi^{(o)\dag}_2\xi_1^{(o)\dag}
+\varphi^\dag\xi^{(o)}_1\xi^{(o)}_2)$ with $g\propto-2U^{(co)}$.

{\it WZ-SUSY Model: Massless. --} For simplicity, we drop the
superscript of $\xi^{(o)}$ hereafter.  By combining (\ref{el}),
(\ref{kg}) and the Yukawa coupling together, the effective
Lagrangian after neglecting $O(|\varphi|^6)$  is given by
\begin{eqnarray}
{\cal L}(\xi,\varphi)&=&-\frac{1}2\partial_\mu
\varphi^\dag\partial^\mu
\varphi-\frac{1}2\varepsilon_m|\varphi|^2-i\xi^\dag\sigma^\mu\partial_\mu\xi\nonumber\\
&-&\frac{\lambda}8|\varphi|^4-\frac{g}2(\varphi\xi^\dag_2\xi_1^\dag+\varphi^\dag\xi_1\xi_2).
\label{massless}
\end{eqnarray}
 Tuning $\varepsilon_m=0$ by varying $U^{(c)}$, and
further tuning pair-pair (or molecule-molecule) interaction by
varying $U^{(co)}$ so that the coupling constant $\lambda=g^2$, the
effective Lagrangian ${\cal L}(\xi,\varphi)$ is exactly the {\it
massless WZ-SUSY model} with the SUSY under the SUSY transformations
$\delta \varphi=\epsilon^\dag\sigma^y\xi,$ and
$\delta\xi=\sigma^\mu\sigma^y\epsilon\partial_\mu\varphi^\dag-\frac{g}2\varphi^{\dag
2}\epsilon$ where $\epsilon$ is a constant two-component spinor
parameter.

{\it WZ-SUSY Model: Massive. --} To have a massive WZ-SUSY model, we
need to introduce an external source. This can be realized by
putting a Bose-Einstein condensate of dimolecules nearby, which is
made of pairs of molecules (or 4-atom molecules) (see Fig. 1(a)).
Through Josephson tunneling with an amplitude $\kappa$, the
dimolecule condensate exchanges pairs of molecules with the mixture.
The effective Lagrangian reads
\begin{eqnarray}
\label{source} {\cal L}(\xi,\varphi,\Psi)&=&-\frac{1}2\partial_\mu
\varphi^\dag\partial^\mu
\varphi-i\xi^\dag\sigma^\mu\partial_\mu\xi\nonumber\\
&-&\frac{g^2}8|\varphi|^4
-\frac{g}2(\varphi\xi^\dag_2\xi_1^\dag+\varphi^\dag\xi_1\xi_2)
\nonumber\\&+&\kappa(\Psi^{\dag}\varphi^2+\Psi\varphi^{\dag2}),
\end{eqnarray}
 where $\Psi$ is the external dimolecular
field. There is a global U(1) symmetry (called R-symmetry) under
$\xi \to e^{i\theta}\xi$, $\varphi\to e^{2i\theta}\varphi$ and
$\Psi\to e^{4i\theta}\Psi$ \cite{rsymm}. If $\Psi$ slowly varies in
space-time, it is also SUSY invariant under $\delta
\varphi=\epsilon^\dag\sigma^y\xi$ and
$\delta\xi=\sigma^\mu\sigma^y\epsilon\partial_\mu\varphi^\dag-\frac{g}2\varphi^{\dag
2}\epsilon+\frac{4\kappa\Psi^{\dag}}{g}\epsilon$. By taking $\Psi$
to be its condensed order parameter
$\langle\Psi\rangle=\langle\Psi^{\dag}\rangle=m^2/8\kappa$, the
R-symmetry is broken and reduced to a discrete $\mathbb{Z}_2$
symmetry with $\xi\to i\xi$ and $\varphi\to-\varphi$, and  the
on-shell WZ Lagrangian appears (up to an additive constant)
\begin{eqnarray}
{\cal L}(\xi,\varphi,m)&&=-\frac{1}2\partial_\mu
\varphi^\dag\partial^\mu
\varphi-i\xi^\dag\sigma^\mu\partial_\mu\xi\\
&&-\frac{g^2}8(\varphi^{\dag2}
-\frac{m^2}{g^2})(\varphi^2-\frac{m^2}
{g^2})\nonumber\\&&-\frac{g}2(\varphi\xi^\dag_2\xi_1^\dag+\varphi^\dag\xi_1\xi_2).\nonumber\label{os}
\end{eqnarray}
The SUSY is exact by replacing $\Psi^{\dag}$ with
$\langle\Psi^{\dag}\rangle$ in the SUSY transformations. The
$\mathbb{Z}_2$ symmetry is always spontaneously broken in one of
the
degenerate ground states with $\varphi=\phi\pm m/g$. The SUSY
Lagrangian with spontaneous breaking of $\mathbb{Z}_2$ becomes
\begin{eqnarray}
{\cal L}&=&-\frac{1}2\partial_\mu \phi^\dag\partial^\mu
\phi-i\xi^\dag\sigma^\mu\partial_\mu\xi
\mp\frac{1}2m(\xi^\dag_2\xi_1^\dag+\xi_1\xi_2)-\frac{g^2}8|\phi|^4\nonumber
\\
&&-\frac{1}2m^2|\phi|^2\mp\frac{gm}4|\phi|^2
(\phi+\phi^\dag)\nonumber
\\
&&-\frac{g}2(\phi\xi^\dag_2\xi_1^\dag
+\phi^\dag\xi_1\xi_2).\label{onshell}
\end{eqnarray}
This is the 2+1-dimensional reduction of the original WZ-SUSY model
in 3+1 dimensions \cite{wz,Weinberg}.

{\it Supercurrent and Supercharge. --} SUSY leads to a conserved
supercurrent, whose conserved supercharges are generators of SUSY.
The supercurrent is defined by $ \delta\int dtd^2x {\cal L}=\int
dtd^2x \epsilon^\dag\sigma^y\partial_\mu J_s^\mu $. The supercharges
are then given by $Q=\int d^2x J_s^0(x)$ and $Q^\dag$. The SUSY
transformation generated by $Q$ for a field $O$ reads $\delta
O=-i\epsilon^\dag\sigma^y [Q,O]_\pm$. We focus on the on-shell model
with the $\mathbb{Z}_2$ symmetry spontaneously broken, where the
on-shell supercurrent\cite{supc} is given by $
J^\mu_s=i\sigma^\mu\sigma^\nu\xi\partial_\nu\phi
+i\frac{g}2(\phi^2\pm2m\phi/g)\sigma^y\sigma^\mu\xi.$ The SUSY
spontaneous breaking is signaled by $\langle\{Q,O\}\rangle\ne 0$ for
a fermionic operator $O$. However, for this simplest SUSY model, the
SUSY is not spontaneously broken at zero temperature
\cite{Weinberg}.

{\it Nonrenormalization . --} In this simplest WZ-SUSY model,
single-particle Green's functions are not renormalized due to
(unbroken) SUSY. For example, the renormalization to Klein-Gordon
field's propagator in a one-loop self-energy calculation is given by
$q^2-m^2\to q^2-m^2_\phi(q)$ with \cite{phonino} $
m_\phi(q)=m+g_R\langle A\rangle_0+O(g_R^2\langle A\rangle_0^2)$
where $A={\rm Re}~\phi$. For 2+1 dimensions, due to the nonzero
anomalous critical exponents \cite{sei}, the coupling constant may
be renormalized to $g_R$. However, $\langle A\rangle_0\propto
\langle\{Q,\xi\}\rangle_0 =0$ because  the SUSY is not spontaneously
broken. The mass of the Dirac field is also not renormalized as
required by SUSY. Therefore, the single particle
 Green's functions, both of the Dirac and
Klein-Gordon fields, are not renormalized from their free version.
The spectral functions of the Green's functions can be measured by
the single-particle spectroscopic technique which has been developed
recently \cite{sps}. The nonrenormalization of the Green's functions
implies sharp peaks in their spectral functions, with identical {\em
relativistic} dispersions for the atoms and molecules.
Experimentally this would be the hallmark of achieving SUSY.

{\it Thermal breaking of SUSY. --} By replacing $t$ by $i\tau$, the
imaginary time, the Euclidean version of Lagrangian (\ref{onshell})
describes WZ-SUSY model in finite temperature $T$. When $T\ne 0$,
SUSY is always broken because $
\langle\{Q,Q^\dag\sigma^y\}\rangle_T=\langle\sigma^\mu
P_\mu\rangle_T\ne 0$ with $P_\mu$ being the energy-momentum operator
\cite{tb}, due to the nonvanishing thermal energy. This SUSY thermal
breaking is accompanied by a thermal Goldstone fermion (phonino) but
not necessarily by a phonon because the Lorentz symmetry is also
broken by $\langle P_0\rangle_T\ne 0$\cite{phonino}. The phonino
dispersion is given by\cite{phonino} $q_0=\pm v_{ss}|\bf q|$ where
the SUSY sound velocity $v_{ss}=v_s/3$ for $T\gg m$ and
$v_{ss}=Tv_s/m$ for $T\ll m$.

To detect the phonino mode, one can consider the response to an
external 'fermionic' field coupled to the supercurrent. The phonino
is a pole of the supercurrent-supercurrent correlation function.
This external 'fermionic' field can be a combination of an external
photon with another hyperfine state of the fermionic atom which is
decoupled to the mixture. We have studied this kind of SUSY response
theory for a nonrelativistic SUSY mixture \cite{SYS}. However, the
difficulty in the present case is that the supercurrent is not so
simple as that in the nonrelativistic theory, and thus the coupling
between the external 'fermionic' field and the supercurrent is not
that easy to be experimentally handled.

Replacing $\langle A\rangle_0$ by $\langle A\rangle_T$, the masses
are thermally renormalized. The masses of $A$, $B$ ($\phi=A+iB$) and
the spinor $\xi$ have been calculated in low temperature and high
temperature limits\cite{phonino}. Namely, in 2+1-dimensions up to
one-loop, for $T\ll m$, one has $m_B=m$, $ m^2_A-m^2_B\propto
g^2m\alpha,m_\xi-m_A\propto g^2m\alpha, $ where
$\alpha=\frac{2T}{\pi m}e^{-m/T}$;  for $T\gg m$, $m_B=m$, $
m^2_A=m^2-2g^2T,~m^2_\xi=m^2-g^2T$. These unequal masses of these
fields signal SUSY breaking, and can be probed quantitatively. In
particular, we spectroscopy measurements to show double peaks in the
molecule spectral function due to the unequal masses  between $A$
and $B$ components, while the atom spectral function has a single
peak with a mass of the Dirac field equal to neither $m_A$ nor
$m_B$.

{\it Experimental Challenges. --} Optical lattices that trap cold
atoms can be routinely set up in laboratory. The staggered Peierls
phase originates from production of the artificial magnetic field
\cite{spielman}. As discussed earlier, one needs to tune three
parameters to achieve SUSY \cite{notes}: atom-atom interaction,
molecule-molecule interaction, and molecule velocity. The former two
may be done by adjusting the real magnetic field in Feshbach
resonance while the latter may be related to the interaction between
laser field and the Feshbach molecule. These are all achievable
within existing experimental capabilities. Perhaps the biggest
challenge is finding the right fermionic atom, which needs to have a
highly tunable interaction through a $p$-wave Feshbach resonance. It
also needs to support a sufficiently stable dimolecule  (or 4-atom
bound) state, whose condensate provides the source term in Eq.
(\ref{source}), which gives rise to equal particle masses.
Experimentally, one needs to overcome the atom loss due to the
heating of the atom gas caused by three- and four-body collisions.
Without the last ingredient however, one can still realize the
massless version of the WZ-SUSY model, Eq. (\ref{massless})(with
$\epsilon_m$ tuned to zero), which already contains very rich SUSY
physics. Despite these and other challenges, we believe simulating
the WZ-SUSY model using cold atom-molecule mixtures is a worthwhile
endeavor, as as generalizing the present model to a
(3+1)-dimensional model is straightforward, and then it provides new
opportunities to explore the real space-time SUSY physics.

YY thanks Q. J. Chen, C. Liu and M. X. Luo for helpful discussions.
This work was supported by NNSF of China, the national program for
basic research of MOST of China, and the Key Lab of Frontiers in
Theor. Phys. of CAS (YY), and by National Science Foundation grant
Nos. DMR-0704133 and DMR-1004545 (KY).


\begin{references}
\bibitem{notes} Sometimes, although the SUSY is explicitly broken,
the SUSY effects can still be observed if the system only deviates
slightly from the SUSY, e.g., the response to the photoassociation
perturbation we studied in \cite{SYS}.
\end{references}
\end{document}